\documentclass[journal]{IEEEtran}

\usepackage[compress]{cite}
\usepackage{stmaryrd}
\usepackage{mathrsfs}
\usepackage{amsfonts}
\usepackage{amsmath}
\usepackage{cases}
\usepackage{txfonts}
\usepackage{amssymb}
\usepackage{epsfig}
\usepackage{pdfpages}
\usepackage{algorithm}
\usepackage{algpseudocode}

\usepackage{multirow}

\begin{document}

	\title{Enhanced ODMA for Massive Sparse Access: Hybrid Data-Frozen Bit Transmission and Fixed-Point Analysis over Block Fading Channels}

	\author{Guanghui Song, \IEEEmembership{Member, IEEE},
		Jianxiang Yan,
		Ying Li, \IEEEmembership{Member, IEEE}, Zhaoji Zhang, \IEEEmembership{Member, IEEE}\\ and
		Jun Cheng, \IEEEmembership{Member, IEEE}
		
		\thanks{
		Guanghui Song, Jianxiang Yan, Ying Li, and Zhaoji Zhang are with School of Communication Engineering, Xidian University, Xi'an, 710071, China (songguanghui@xidian.edu.cn, jianxiangyan@stu.xidian.edu.cn, yli@mail.xidian.edu.cn, zhaojizhang@xidian.edu.cn). 
				
Jun Cheng is with the Department of Intelligent Information Engineering
and Science, Doshisha University, Kyoto 610-0321, Japan (jcheng@ieee.org).
	}	
	}

	\maketitle

	\begin{abstract}
This paper proposes a novel on-off division multiple access (ODMA) transmission scheme that enables efficient joint multi-user channel estimation and iterative decoding by inserting a small number of frozen symbols into the codewords. Functionally analogous to pilots, these symbols are sparsely distributed within the codeword. Unlike conventional pilot-based methods, our approach requires only a minimal number of frozen symbols (e.g., $5\sim20$ symbols per user in a 300-user system), which serve dual purposes as both estimation references and decoding aids. By employing low-complexity single-user channel estimation and decoding, combined with simple iterative interference cancellation, the scheme achieves performance equivalent to that with perfectly known user channels, even when accounting for the additional energy and bandwidth costs of the frozen symbols. Furthermore, for the large-scale ODMA sparse multiple access system, this paper proposes a fixed-point analysis method, which can accurately estimate the iterative convergence performance over multi-user block fading channels by only leveraging the decoding functions under the single-user AWGN channel. This method is applicable to performance analysis for arbitrary code lengths, code rates, and decoding algorithms. It eliminates the need for extensive Monte Carlo simulation time, and provides an efficient tool for the design of multi-user codes.
	\end{abstract}
	
	
	\section{Introduction}
	\IEEEPARstart{I}{n} the context of grant-free non-orthogonal multiple access (NOMA), a massive number of Internet of Things (IoT) devices access the network randomly in bursty, short-packet transmissions, imposing stringent requirements for low latency and high connection density. The conventional approach of allocating long pilot sequences to each user consumes substantial wireless resources, severely limiting access capacity and transmission efficiency. Therefore, it is imperative to reduce pilot length to minimize overhead. However, extremely short pilots lead to a sharp decline in channel estimation accuracy. If estimation and decoding are processed separately, system performance will degrade significantly. Consequently, the adoption of joint channel estimation and decoding becomes essential. By utilizing all users' data symbols as ``virtual pilots" during the iterative decoding process to dynamically refine channel information, reliable user detection and data recovery can be achieved with very low pilot overhead. This is a necessary technical pathway to realize the coexistence of massive connectivity and high spectral efficiency.
	
	Conventional joint channel estimation and signal detection schemes are typically characterized by the use of long pilots or spreading sequences, combined with computationally intensive multi-user linear minimum mean square error (LMMSE) estimation or factor graph processing \cite{CS1, KG3CS, KG4CS, KG5AMP, KG6OMP,KG1,KG2GAMP,BDF2,BC1,ZZJ1,ZZJ2,BC3,BD3,BD4,YW,ZYY,BDDP}. A common limitation of these approaches is their inadequate consideration of channel coding.
Recently, we proposed a transmission scheme that operates without pilots or spreading sequences and performs multi-user channel estimation aided by channel decoding \cite{Yifei}.  In the absence of dedicated pilots or spreading, each user's signal is marked by a unique scrambling pattern that serves as a user-specific signature. The receiver utilizes this scrambling information to perform a single-user maximum-likelihood channel estimation, and then iterates between the decoder and the channel estimator to progressively improve the accuracy of both channel estimates and decoded data.
However, two main limitations remain in \cite{Yifei}. First, estimating each user's channel requires solving a fixed-point equation array, which entails high computational complexity. Second, the method requires long data frames to achieve accurate channel estimation, and its performance degrades severely when the number of users is large, making it unsuitable for massive-user short-packet communication scenarios.

Another challenging issue is that most existing performance analysis works \cite{NgoA,ChenC,RaviT,Kowshik,Gao1,Polyanskiy,LangY,Kowshik2,Gao2,Fengler,Yavas,Yavas2,Lancho,Gao3} focus on random codes and maximum-likelihood-like decoding. The resulting performance bounds offer limited guidance for practical system design. 
Traditional density evolution and extrinsic information transfer (EXIT) chart methods are both based on the assumptions of infinite code length and ergodic channels, which are not applicable to short-data coding and block fading channel scenarios. Consequently, the performance prediction of various schemes usually relies on extensive Monte Carlo simulations, which becomes computationally prohibitive under massive user access conditions.

This paper first presents a novel multi-user transmission scheme to overcome the aforementioned limitations in \cite{Yifei}. By adopting the on-off division multiple access (ODMA) sparse transmission framework \cite{ODMA,Ozates,Hao,ZZhang,Yanlett,Yantcom,ZZhangIot,Survey,Gao}, our scheme efficiently supports massive-user short-packet communications. A key feature is the insertion of very few frozen symbols (e.g., $5\sim20$ per user for 300 users) into each codeword to assist channel estimation, eliminating complex signal scrambling. Channel estimation is performed via low-complexity univariate  LMMSE, which involves only two vector correlation operations, circumventing high-complexity fixed-point solutions. Simulation results show that the proposed scheme, equipped with low-complexity iterative interference cancellation, achieves performance comparable to the perfect channel state information (CSI) benchmark for 300 users, provided that the energy overhead incurred by frozen symbols is taken into consideration.

Another contribution of this paper lies in the proposed fixed-point analytical framework for multi-user iterative decoding, which accommodates massive sparse access, block-fading channels, and short-packet communication scenarios.
Given the time-domain non-ergodic property of block-fading channels, we leverage the user-domain ergodicity brought by a large number of users to characterize the average decoding performance across all users.
To tackle the performance evaluation difficulties for short-packet coding, we map the multi-user iterative decoding process to a large factor graph, where each single-user decoder acts as a processing node.
Within this framework, only the input-output transfer function of the single-user decoder needs to be precomputed via Monte Carlo simulations. The overall iterative convergence performance of multi-user detection and decoding can then be readily derived by solving the corresponding fixed-point equation.
We further validate the fixed-point performance for various channel codes with distinct code rates. Numerical results show that the optimal achievable performance is merely 1.5 dB away from the corresponding converse bound.

\section{Transmission Scheme}
\begin{figure}[t]
	\includegraphics[width=
	3.5 in]{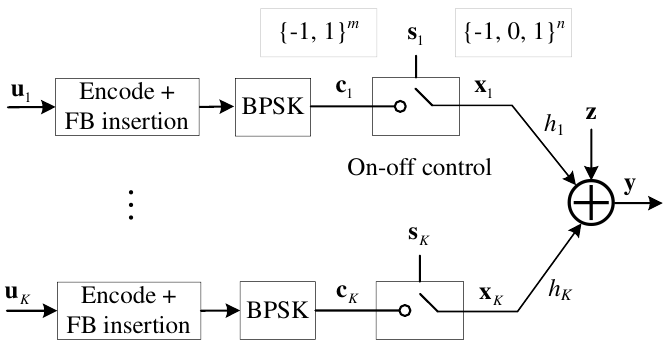}
	\centering
	\caption{Hybrid data and frozen bit (FB) transmission with an ODMA mechanism.} \label{fig:system}
\end{figure}

We consider a $K$-user multiple access communication scenario as shown in Fig.~\ref{fig:system}. For the $k$-th user, the information data $\mathbf{u}_k$ (of length $B$) first undergoes channel coding, yielding a codeword of length $L$. Then, frozen bits of length $P$ are sparsely inserted into the codeword. The insertion positions of the frozen bits are unique for each user. For simplicity, all frozen bits are set to 0 in our transmission scheme. These frozen bits aid in multi-user channel estimation and decoding. The data is modulated and transmitted in an ODMA manner. Here, for simplicity, we take BPSK modulation $(0\rightarrow 1, 1\rightarrow -1)$ as an example; however, our work is applicable to general modulation schemes.

ODMA is a sparse access scheme \cite{ODMA} that departs from the conventional spreading framework, primarily leveraging channel codes and users' sparse access patterns for joint multi-user iterative detection. Taking the processing of the $k$-th user as an example, let $\mathbf{c}_k=(c_{k,1}, c_{k,2},...,c_{k,m})$ denote the modulated output of the $k$-th user, which has a length of $m=L+P$. These 
$m$ symbols are transmitted over 
$n$ channel uses, where 
$n>m$. That is, only 
$m$ of the 
$n$ channel uses actually carry data, while the remaining 
$n-m$ slots are idle. The specific channel uses used for transmission are determined by user $k$'s on-off pattern 
$\mathbf{s}_k=(s_{k,1}, s_{k,2},...,s_{k,n}), s_{k,i}\in\{0,1\}$ for $i=1,2,...,n$. Here, 
$\mathbf{s}_k$ is an 
$n$-long vector with a weight of 
$m$. The 
$m$ non-zero positions indicate the channel uses in which data is transmitted.  A design principle of the on-off pattern is to make the number of accessing users in each time slot as equal as possible and sufficiently random (the on-off pattern matrix avoids short cycles). The transmitted signal $\mathbf{x}_k=(x_{k,1}, x_{k,2},...,x_{k,n})$
of the $k$-th user is expressed as
\begin{equation}
x_{k,i}=\begin{cases}
0, & \text{if } s_{k,i} = 0 \\
c_{k,w_i}, & \text{if } s_{k,i} = 1
\end{cases}\label{eq:xc}
\end{equation}
where $w_i=\sum_{j=1}^is_{k,j}$.

\begin{figure}[t]
	\includegraphics[width=
	3.3 in]{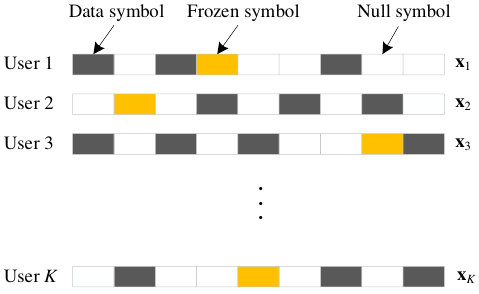}
	\centering
	\caption{Codeword structures for $K$-user ODMA transmission.} \label{fig:signal}
\end{figure}

The resulting codeword transmission structure for the $K$ users is illustrated in Fig.~\ref{fig:signal}. As can be observed, a key distinguishing feature of our scheme compared to conventional communication systems is the inclusion of three distinct types of symbols in the codeword: data symbols, frozen symbols, and null symbols (which correspond to the idle slots in ODMA). The unique arrangement of these three symbol types for each user forms a distinctive signature. This codeword structure is particularly advantageous for multi-user channel estimation and iterative decoding.

Similar to the original ODMA scheme \cite{ODMA}, we adopt non-random multiple access, where each user's on-off pattern and frozen-bit insertion pattern are pre-assigned and known to the receiver. By contrast, in the unsourced random access (URA) scenario \cite{Ozates,Hao,ZZhang,Yanlett,Yantcom,ZZhangIot,Survey,Gao}, such patterns are inherently determined by the data transmitted by each user. By enabling these patterns to be selected dynamically from transmitted data, the proposed scheme can be readily extended to URA scenarios.

We consider a quasi-static Rayleigh fading channel where the fading coefficient $h_k\sim \mathcal{CN}(0,1)$ remains constant over $n$ channel uses and independently changes every $n$ channel uses. Thus, the expression for the received signal is
\begin{equation*}
\mathbf{y}=\sum_{k=1}^Kh_k\mathbf{x}_k+\mathbf{z}
\end{equation*}
where $\mathbf{z} = (z_1, z_2, ..., z_n)$, with $z_i\sim \mathcal{CN}(0, N_0)$, is an independent and identically distributed (i.i.d.) Gaussian noise vector.

\section{Joint Multi-User Channel Estimation and Decoding}

\begin{figure}[t]
	\includegraphics[width=
	3.0 in]{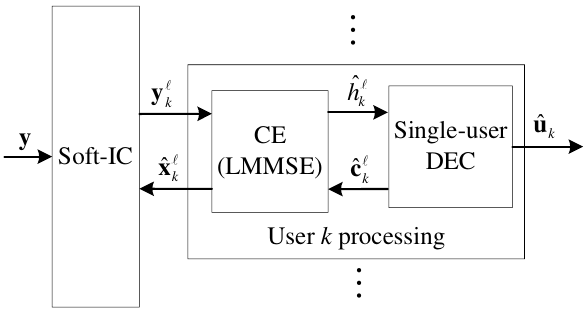}
	\centering
	\caption{A soft interference cancellation (Soft-IC)-based iterative receiver for joint multi-user channel estimation (CE) and decoding (DEC).} \label{fig:receiver}
\end{figure}

We propose an efficient joint channel estimation and decoding method for $K$ users, whose complexity is approximately $K$ times that of single-user decoding. The overall receiver framework is illustrated in Fig.~\ref{fig:receiver}. In general, joint multi-user channel estimation and decoding is implemented in an iterative manner. Each user's channel estimation and decoding can be processed independently and in parallel, followed by a simple soft interference cancellation to progressively eliminate inter-user interference. The goal of joint multi-user decoding is gradually achieved through multiple iterations. Furthermore, channel estimation utilizes the entire received codeword (including both frozen symbols and unknown data symbols). In each iteration, the DEC module outputs an estimate of the data symbols to the CE module. The CE module then updates the channel information based on this estimate. We propose a low-complexity univariate LMMSE channel estimation method that requires only two simple correlation operations to implement.

\subsection{Soft Interference Cancellation}
We consider the processing in the $\ell$-th iteration. Let $\hat{h}_j^{\ell-1}$ and $\hat{\mathbf{x}}_j^{\ell-1}$ denote the estimated channel coefficient and codeword of user $j$ obtained in the $(\ell-1)$-th iteration. Initially, we have $\hat{h}_j^0=0$. The estimates for all data bits are initialized to $0$, implying that $+1$ and $-1$ are considered equally likely. Null symbols and frozen symbols are initialized to their actual transmitted values: null symbols are 0, and frozen symbols are 1. The signal output by the soft interference canceller to the processing module of the $k$-th user is the residual signal after subtracting the estimates of the other users' signals from the received signal $\mathbf{y}$, denoted as
\begin{equation}
\mathbf{y}_k^{\ell}=\mathbf{y}-\sum_{j\neq k}\hat{h}_j^{\ell-1}\hat{\mathbf{x}}_j^{\ell-1}.\nonumber
\end{equation}
After obtaining $\mathbf{y}_k^{\ell}$ for $k=1,2,\ldots,K$, the CE and DEC modules perform channel estimation and decoding for each user in parallel.
\subsection{Channel Estimation}
We consider the processing of user $k$. First, $\mathbf{y}_k^{\ell}$ can be rewritten as 
\begin{align}
\mathbf{y}_k^{\ell}=&h_k\mathbf{x}_k+\mathbf{z}_k^\ell\label{eq:yk}\\
&\mathbf{z}_k^\ell=\sum_{j\neq k}\Big(h_j\mathbf{x}_j-\hat{h}_j^{\ell-1}\hat{\mathbf{x}}_j^{\ell-1}\Big)+\mathbf{z}\nonumber
\end{align}
where $\mathbf{z}_k^\ell$  is the residual interference.

We treat $\mathbf{z}_k^\ell$ as a complex Gaussian vector characterized by entries $z_{k,i}^\ell \sim \mathcal{CN}(0, V_{k,i}^\ell)$, with $V_{k,i}^\ell$ denoting the residual interference variance of user $k$ at channel use $i$. 
 Since we have obtained an estimate $\hat{h}_j^{\ell-1}$ of $h_j$, $h_j$ can be modeled as a complex Gaussian distribution $h_j \sim \mathcal{CN}(\hat{h}_j^{\ell-1}, e^{\ell-1}_j)$, where $e^{\ell-1}_j=E[|h_j-\hat{h}_j^{\ell-1}|^2]$ is the MSE of the channel estimation of user $j$ in the $(\ell-1)$-th iteration. Each $\mathbf{x}_j$ is treated as a Bernoulli
 vector with mean $\hat{\mathbf{x}}_j^{\ell-1}$. Thus, the residual interference variance is derived as
 \begin{align}
 V_{k,i}^\ell&=\text{Var}\big[z_{k,i}^\ell\big]\nonumber\\
 &=\sum_{j\neq k}\text{Var}\big[h_jx_{j,i}\big]+N_0\nonumber\\
 &=\sum_{j\neq k}E\Big[\big|h_jx_{j,i}\big|^2\Big]-\big|E[h_jx_{j,i}]\big|^2+N_0\nonumber\\
 &=\sum_{j\neq k}s_{j,i}\Big(\big|\hat{h}_j^{\ell-1}\big|^2+e^{\ell-1}_j-\big|\hat{h}_j^{\ell-1}\big|^2(\hat{x}_{j,i}^{\ell-1})^2\Big)+N_0.\label{eq:Vk}
 \end{align}

Based on \eqref{eq:yk}, the LMMSE estimate of $h_k$ and the corresponding MSE $e^{\ell}_k$ are derived in Appendix~\ref{app:LMMSE}, with the results given as
\begin{align} 
\hat{h}_k^\ell &= \frac{1}{ 1 + \sum_{i=1}^ns_{k,i} \frac{(\hat{x}^{\ell-1}_{k,i})^2}{1-(\hat{x}^{\ell-1}_{k,i})^2+V_{k,i}^\ell} } \sum_{i=1}^ns_{k,i} \frac{\hat{x}^{\ell-1}_{k,i} y^{\ell}_{k,i}}{1-(\hat{x}^{\ell-1}_{k,i})^2+V_{k,i}^\ell}\label{eq:CE}\\
e^{\ell}_k &=E\Big[\big|h_k-\hat{h}_k^\ell\big|^2\Big]=\frac{1}{ 1 + \sum_{i=1}^ns_{k,i} \frac{(\hat{x}^{\ell-1}_{k,i})^2}{1-(\hat{x}^{\ell-1}_{k,i})^2+V_{k,i}^\ell} }.\label{eq:MSE}
\end{align}
Since the LMMSE estimation here is for a single variable, it involves only two simple vector correlation operations and does not include complex matrix operations.

It should be emphasized that without frozen bits, we would have $\hat{\mathbf{x}}_j^{\ell-1}=\textbf{0}$ at the initial iteration $\ell=1$, which would lead to $\hat{h}_k^\ell=0$ and cause the iterative decoding to stall. Therefore, frozen bits can serve as a small number of pilots to enable the iteration to start at the initial decoding stage. Consequently, frozen bits are a prerequisite for employing LMMSE estimation here, thus avoiding the high-complexity fixed-point solving operations required by ML estimation as in\cite{Yifei}.

Although the frozen bits in our scheme can serve the role of pilots in traditional communication systems, there are significant differences between the two. Traditional schemes rely on pilots for channel estimation, which are typically inserted at the head of the codeword. The subsequent data portion does not participate in channel estimation. When the number of users is large, the required pilots are usually long. In contrast, our scheme utilizes the entire received signal for joint channel estimation and decoding, where the frozen bits play an auxiliary role and their length can be flexibly controlled.
 
 \subsection{Single-User Decoding}
 Single-user decoding typically adopts a soft-input soft-output decoding algorithm, such as the belief propagation (BP) algorithm, which is widely used for decoding high-performance modern channel codes like low-density parity-check (LDPC) codes. The input and output of this decoder are typically represented by the log-likelihood ratio (LLR) of the coded bits. Still taking the decoding of the $k$-th user as an example, first, the DEC discards the received signals corresponding to frozen symbols and null symbols based on user $k$'s frozen-bit insertion pattern and on-off pattern, and performs decoding using only the residual signals of the data symbols.
 \subsubsection{Input LLR}
The DEC for user $k$ should first compute an initial LLR for each coded bit in $\mathbf{c}_k$ using $ \hat{h}_k^\ell$ and $\mathbf{y}_k^{\ell}$. Unlike the conventional LLR calculation with perfect channel knowledge, here the channel coefficient $h_k$ is unknown. Since we have obtained an estimate $\hat{h}_k^\ell$ of $h_k$, $h_k$ can be modeled as a complex Gaussian distribution $h_k \sim \mathcal{CN}(\hat{h}_k^\ell, e^{\ell}_k)$.
According to \eqref{eq:xc}, for $s_{k,i}\neq0$, $c_{k,w_i}$ (where $w_i = \sum_{j=1}^i s_{k,j}$) is related to $x_{k,i}$ and consequently to $y^{\ell}_{k,i}$. Thus, for $s_{k,i} \neq 0$, if $c_{k,w_i}$ is a data symbol, its LLR can be calculated as
\begin{align}
L^{\ell}_\text{in}(c_{k,w_i})&=\log\frac{\text{Pr}\Big(c_{k,w_i}=1|y^{\ell}_{k,i},\hat{h}_k^\ell\Big)}{\text{Pr}\Big(c_{k,w_i}=-1|y^{\ell}_{k,i},\hat{h}_k^\ell\Big)}\nonumber\\
&=\log\frac{\text{Pr}\Big(x_{k,i}=1|y^{\ell}_{k,i},\hat{h}_k^\ell\Big)}{\text{Pr}\Big(x_{k,i}=-1|y^{\ell}_{k,i},\hat{h}_k^\ell\Big)}\nonumber\\
&=\frac{4\Re\left((\hat{h}_k^\ell)^*\mathbf{y}_k^{\ell}\right)}{V_{k,i}^\ell+e^{\ell}_k}\label{eq:LLR}
\end{align}
where \eqref{eq:LLR} is derived in Appendix~\ref{app:LLR}.  The difference between \eqref{eq:LLR} and the conventional LLR formula is the additional inclusion of the MSE of the channel estimation in the denominator. Notably, \eqref{eq:LLR} implies that the signal-to-noise ratio (SNR) for user $k$ is
\begin{eqnarray}
 \text{SNR}_k=\frac{2\Re\left((\hat{h}_k^\ell)^*h_k\right)}{V_{k,i}^\ell+e^{\ell}_k}\label{eq:snr}
\end{eqnarray}
 after performing Soft-IC. 

\subsubsection{Formulate $\hat{\mathbf{x}}_k^{\ell}$}
Using $L^{\ell}_\text{in}(\mathbf{c}_k)$, the DEC of user $k$ performs a soft decoding and outputs a more reliable LLR 
 \[L^{\ell}_\text{out}(c_{k,i})=\log\frac{\text{Pr}(c_{k,i}=1|\mathbf{y}_k^{\ell}, \hat{h}_k^\ell )}{\text{Pr}(c_{k,i}=-1|\mathbf{y}_k^{\ell}, \hat{h}_k^\ell )}\]
 for each data symbol in $\mathbf{c}_k$.  An estimate of $c_{k,i}$ can be obtained as 
  \begin{equation}
  \hat{c}_{k,i}=E\Big[c_{k,i}|\mathbf{y}_k^{\ell}, \hat{h}_k^\ell\Big]=\tanh\left(\frac{L^{\ell}_\text{out}(c_{k,i})}{2}\right).\nonumber
  \end{equation}
   The decoding output squared error is represented by 
  \begin{equation}
  e^\ell_{k,i}=\big|c_{k,i}-\hat{c}_{k,i}^{\ell}\big|^2.\label{eq:mse}
  \end{equation}
If $x_{k,i}$ is a data symbol, its estimate can be derived as 
  \begin{equation}
\hat{x}_{k,i}^\ell=\hat{c}_{k,w_i}^\ell
\end{equation}
where $w_i=\sum_{j=1}^is_{k,j}$.
On the other hand, if $x_{k,i}$ is a frozen or null symbol, we directly set $\hat{x}_{k,i} = x_{k,i}$.

\section{Fixed-Point Analysis over Block Fading Channels}
\begin{figure}[t]
	\includegraphics[width=
	2.0 in]{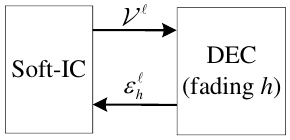}
	\centering
	\caption{Variance evolution of iterative multi-user decoding.} \label{fig:analysis}
\end{figure}

In this section, we analyze the performance of multi-user iterative decoding under the condition of perfectly known channels. This performance serves as a benchmark for our joint multi-user channel estimation and decoding scheme. Since quasi-static fading channels are non-ergodic in the time dimension, the decoding performance in each instance depends on the multi-user channel coefficients. Consequently, their performance cannot be characterized using conventional analytical methods designed for ergodic channels. To address this challenge, we consider asymptotic performance analysis in the large-system regime (where the number of users $K$ is large). This setup is also consistent with most existing research works \cite{Ozates,Hao,ZZhang,Yanlett,Yantcom,ZZhangIot,Yifei}. Under such conditions, channel fading exhibits ergodicity across the multi-user dimension.
Furthermore, we will consider the performance analysis of multi-user decoding in the case of short data encoding (with finite $L$). Specifically, we treat the single-user decoder as a processing node in the large-system iterative decoding process. The characteristic function of this node, which depicts the relationship between the output MSE of the single-user decoder and the input SNR, can be obtained via Monte Carlo simulation. Using this characteristic function, we can obtain the convergence performance of multi-user iterative decoding by an elegant fixed-point analysis.

  \begin{figure*}[t]
	\includegraphics[width=
	7.0 in]{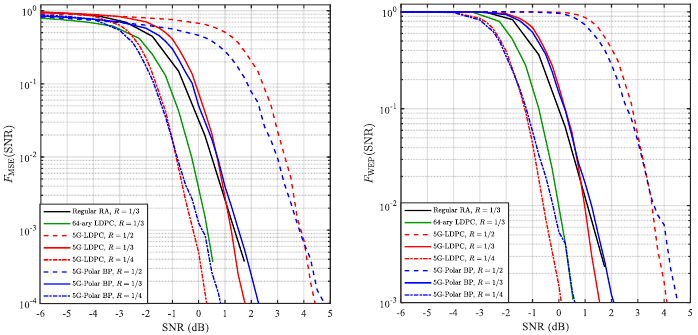}
	\centering
	\caption{Transfer characteristic functions $F_\text{MSE}(\cdot)$ (left) and $F_\text{WEP}(\cdot)$ (right) of channel codes with different code rates $R$. The information bit length is 100 bits. All codes are decoded via the BP algorithm with 100 iterations to guarantee convergence. Numerical results are generated by Monte Carlo simulations under the AWGN channel.} \label{fig:F}
\end{figure*}

\subsection{Variance Evolution Analysis}
Let $f_{k,i}\in\{0,1\}$ be the indicator of data symbols, where $f_{k,i}=1$ indicates that the $i$-th symbol of user $k$ belongs to data symbols, and $f_{k,i}=0$ represents a frozen or null symbol.
We adopt a regular access structure such that every channel use bears the same number of data symbols. We define the access degree $\beta=\sum_{k=1}^K f_{k,i}=KL/n$, which equals the total number of data symbols from all $K$ users mapped to each channel use. This parameter is crucial for characterizing the convergence performance of multi-user iterative decoding.

Since the decoding performance of each user depends on its respective channel fading gain, we examine the decoding convergence behavior for a user with a given channel fading coefficient of 
$h$. The convergence behavior is subsequently established by tracking the evolution of the residual variance $\mathcal{V}^\ell=E[V_{k,i}^\ell]$ from the Soft-IC output and the MSE $\varepsilon^\ell_h=E[e^\ell_{k,i}]=1-(\hat{c}_{k,w_i}^{\ell})^2$ (when $f_{k,i}=1$) of the decoded outputs throughout the iterative multi-user process, as illustrated in Fig.~\ref{fig:analysis}. It is noteworthy that since the residual variance \eqref{eq:Vk} of the Soft-IC output to the $k$-th user does not depend on the channel fading of user $k$, $\mathcal{V}^\ell$ is independent of $h$. However, the MSE of the decoded output $\varepsilon^\ell_h$ is related to the channel $h$ of the user.

\subsubsection{Evolution of $\mathcal{V}^\ell$}
Using \eqref{eq:Vk} and assuming perfect channel state information, i.e., $\hat{h}_j^{\ell-1}=h_j,e^{\ell-1}_j=0$, the average residual variance is derived by  
 \begin{align}
\mathcal{V}^\ell&=\sum_{j\neq k}s_{j,i}E\Big[\Big(\big|h_j\big|^2-\big|h_j\big|^2(\hat{x}_{j,i}^{\ell-1})^2\Big)\Big]+N_0 \nonumber\\
&=\sum_{j\neq k}f_{j,i}E\Big[\big|h_j\big|^2\Big(1-(\hat{c}_{j,w_i}^{\ell-1})^2\Big)\Big]+N_0 \nonumber\\
&=(\beta-1) E\Big[\big|h_j\big|^2\Big(1-(\hat{c}_{j,w_i}^{\ell-1})^2\Big)\Big] +N_0\label{eq:beta}\\
&= (\beta-1) E\Big[\big|h\big|^2\varepsilon^{\ell-1}_h\Big]+N_0\label{eq:EV}
\end{align}
where the expectation $E[\cdot]$ in \eqref{eq:EV} is taken with respect to $h$. Eq.~\eqref{eq:beta} holds because the access degree equals $\beta$, which indicates that each user suffers from $\beta-1$ effective interfering signals transmitted by other users. In addition, all user channels and decoding MSEs are identically distributed, as an identical channel code is adopted for every user.

\subsubsection{Evolution of $\varepsilon^{\ell}_h$}
Let $F_\text{MSE}(\cdot)$ denote the function characterizing the input-output mapping from SNR to decoding MSE for single-user decoding, and we write $\text{MSE} = F_\text{MSE}(\text{SNR})$ to denote this relationship.
For long codewords, this characteristic function can be derived analytically, whereas for short codewords, it is much simpler to acquire via Monte Carlo simulations. Given the residual variance $\mathcal{V}^\ell$ from the Soft-IC output,  the single-user SNR at the output of the soft-IC is $2|h|^2/\mathcal{V}^\ell$, where $h$ is the channel fading coefficient of the user. Thus, the output MSE after single-user DEC is
\begin{equation}
\varepsilon^{\ell}_h=F_\text{MSE}\left(\frac{2\big|h\big|^2}{\mathcal{V}^\ell}\right)\label{eq:F}.
\end{equation}

 \subsubsection{Fixed-Point Equation}
 Combining \eqref{eq:EV} and \eqref{eq:F}, we obtain the fixed-point equation that characterizes the iterative convergence as the number of iterations approaches infinity
  \begin{equation}
  \varepsilon_h=F_\text{MSE}\left(\frac{2\big|h\big|^2}{(\beta-1) E\Big[\big|h\big|^2\varepsilon_h\Big]+N_0}\right)\label{eq:FP}.
  \end{equation}
Equation \eqref{eq:FP} establishes the fixed-point equation for the decoding output MSE $\varepsilon_h$ of a user under a given channel fading coefficient $h$.
The converged MSE of iterative decoding for the user with fading $h$ is equal to the maximum solution of this fixed-point equation within the range $(0,1)$.

   \subsection{Convergence of Average MSE and PUPE}
  Intuitively, evaluating the convergence performance of iterative decoding would require solving the fixed-point equation \eqref{eq:FP} separately for every channel realization $h$ and subsequently averaging the results. This procedure entails solving an infinite set of equations and thus introduces prohibitively high computational complexity. In this section, we present an elegant computational method that requires solving only a single fixed-point equation to obtain the average MSE and per-user probability of error (PUPE) at convergence. 
   
  Specifically, multiplying both sides of \eqref{eq:FP} by $\big|h\big|^2$ and then taking the expectation over $h$, we obtain an alternative fixed-point equation for $\tilde{\varepsilon}=E\Big[\big|h\big|^2\varepsilon_h\Big]$, given by
  \begin{equation}
\tilde{\varepsilon}=E\left[\big|h\big|^2F_\text{MSE}\left(\frac{2\big|h\big|^2}{(\beta-1) \tilde{\varepsilon}+N_0}\right)\right]\label{eq:FP2}.
\end{equation}
Since $\big|h\big|^2$ follows an exponential distribution with parameter 1, \eqref{eq:FP2} can be rewritten as 
  \begin{equation}
\tilde{\varepsilon}= \int_0^{+\infty} te^{-t}F_\text{MSE}\left(\frac{2t}{(\beta-1) \tilde{\varepsilon}+N_0}\right)dt  \label{eq:FP3}.
\end{equation}
Let $\tilde{\varepsilon}^\infty$, $\varepsilon_h^\infty$, and $\varepsilon^\infty$ denote the converged values of $\tilde{\varepsilon}$, $\varepsilon_h$, and $\varepsilon$, as the decoding iteration tends to infinity. We first solve the integral equation \eqref{eq:FP3} on the interval $(0,1)$ to acquire the converged value $\tilde{\varepsilon}^\infty$. 

\subsubsection{Converged MSE}
We then substitute this result into \eqref{eq:FP} to calculate 
  \begin{equation}
\varepsilon_h^\infty=F_\text{MSE}\left(\frac{2\big|h\big|^2}{(\beta-1) \tilde{\varepsilon}^\infty+N_0}\right)\label{eq:ehinf}.
\end{equation}
Taking the expectation of both sides of \eqref{eq:ehinf} over $h$ yields the average output MSE of the iterative decoding
  \begin{align}
\varepsilon^\infty&=E[\varepsilon_h^\infty]\nonumber\\
&=E\left[F_\text{MSE}\left(\frac{2\big|h\big|^2}{(\beta-1) \tilde{\varepsilon}^\infty+N_0}\right)\right]\nonumber\\
&=\int_0^{+\infty} e^{-t}F_\text{MSE}\left(\frac{2t}{(\beta-1) \tilde{\varepsilon}^\infty+N_0}\right)dt  \label{eq:final}.
\end{align}
Note that for the above computation, we only need to solve the fixed-point equation \eqref{eq:FP3} once to obtain $\tilde{\varepsilon}^\infty$.

 Furthermore, when the single-user decoder employs hard-decision decoding, the resultant MSE can be larger than one. Under such circumstances, the solution to \eqref{eq:FP3} should be searched over an extended interval.

\subsubsection{Converged PUPE}
If the decoder output is measured by the word error probability (WEP), we can obtain another transfer characteristic function $\text{WEP}=F_\text{WEP}(\text{SNR})$
for the decoder node, which can be obtained by simulating the frame error rate (FER) of the single-user channel decoder over the AWGN channel. As an example, Fig.~\ref{fig:F} plots the two transfer characteristic functions of various channel codes employed in prior studies \cite{ODMA,Ozates,Hao,ZZhang,Yanlett,Yantcom,ZZhangIot,Survey,Gao}. Each code has an information block length of 100 bits. BP decoding is employed for all codes to produce soft extrinsic information, which facilitates iterative multi-user detection.
Once $\tilde{\varepsilon}^\infty$ is solved, the converged PUPE can be directly derived via the function $F_\text{WEP}(\cdot)$ as follows:
\begin{align}
\text{PUPE}&=E\left[F_\text{WEP}\left(\frac{2\big|h\big|^2}{(\beta-1) \tilde{\varepsilon}^\infty+N_0}\right)\right]\nonumber\\ 
&=\int_0^{+\infty} e^{-t}F_\text{WEP}\left(\frac{2t}{(\beta-1) \tilde{\varepsilon}^\infty+N_0}\right)dt   \label{eq:PUPE}.
\end{align}

Similarly, we may substitute the output metric of the single-user channel decoder in \eqref{eq:PUPE} with bit error probability to compute the average bit error probability of multi-user iterative decoding.

\subsubsection{Approximation Using Discrete Characteristic Functions}
In practice, we may only obtain discrete sampling points of $F_\text{MSE}(\cdot)$ via numerical simulations, denoted as $F_\text{MSE}(t_i)$ for $i=0,1,\dots,T$, where $0\le t_0<t_1<\cdots<t_T<+\infty$. Define $a=2/\left[(\beta-1)\tilde{\varepsilon}+N_0\right]$ and $g(t)=te^{-t}$. The integral in \eqref{eq:FP3} can be numerically approximated as
\begin{align}
&\int_0^{+\infty} g(t)F_\text{MSE}\left(at\right)dt \nonumber\\
= &\frac{1}{a}\int_0^{+\infty} g\left(\frac{t}{a}\right)F_\text{MSE}\left(t\right)dt \nonumber\\
\approx& \frac{1}{a}\sum_{i=1}^T \frac{g\left(\frac{t_i}{a}\right)F_\text{MSE}(t_i)+g\left(\frac{t_{i-1}}{a}\right)F_\text{MSE}(t_{i-1})}{2}\left(t_i-t_{i-1}\right).\nonumber
\end{align}
The integrals in \eqref{eq:final} and \eqref{eq:PUPE} can be computed using the same approximation strategy.

\section{Numerical Results}

\begin{figure}[t]
	\includegraphics[width=
	3.5 in]{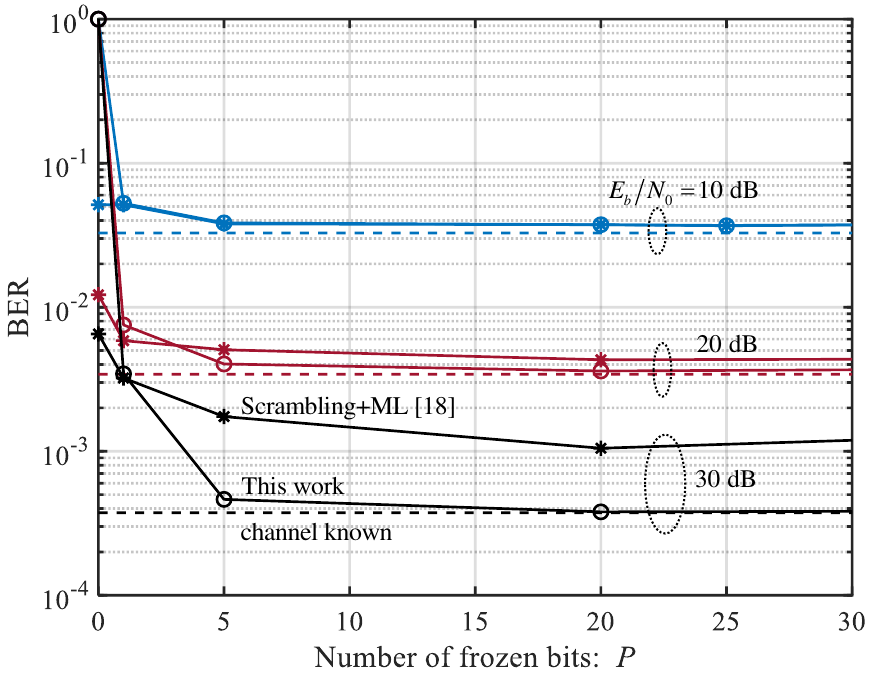}
	\centering
	\caption{Comparison of BER performance over quasi-static Rayleigh fading channels among the scrambling+ML scheme \cite{Yifei}, the case with perfectly known channels, and the proposed scheme in this work. All three cases adopt the same ODMA transmission with the number of users $K=300$, information data length $B=100$, transmission length $n=30000$, and rate-$1/3$ regular-RA code for each user. The maximum number of global iterations for multi-user channel estimation and decoding is set to 30, and each single-user decoding component operates without inner iterations.} \label{fig:BER_frozen}
\end{figure}

\begin{figure}[t]
	\includegraphics[width=
	3.5 in]{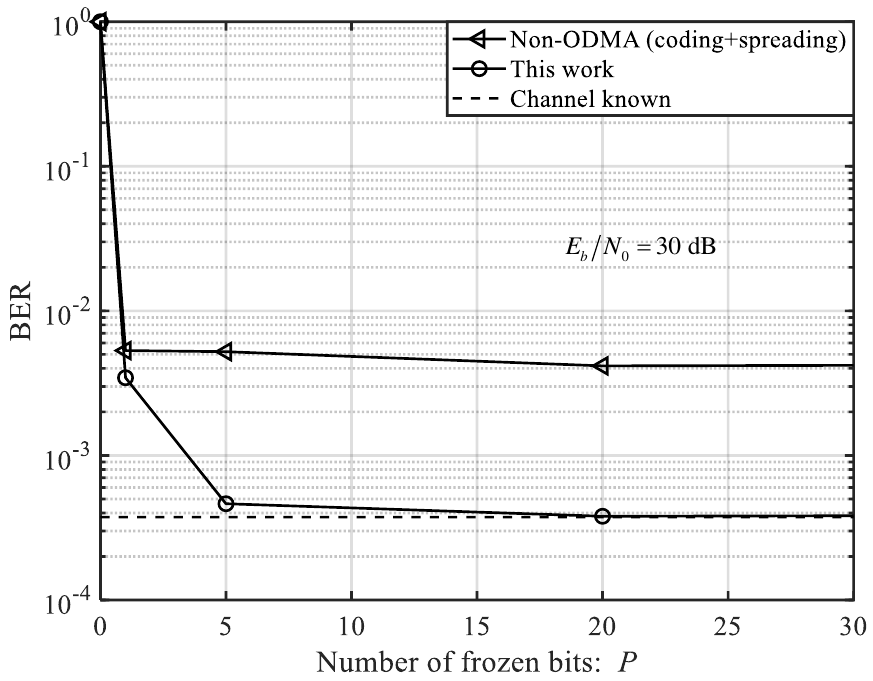}
	\centering
	\caption{Comparison of BER performance between ODMA and non-ODMA (coded-spreading) schemes over quasi-static Rayleigh fading channels with the same receiver algorithm. User number $K = 300$, information data length
		$B = 100$, transmission length $n = 30000$, and rate-$1/3$ regular-RA code for
		each user. The maximum number of global iterations for multi-user channel estimation and decoding is set to 30, and each single-user decoding component operates without inner iterations.} \label{fig:BER_nonODMA}
\end{figure}

In this section, we evaluate the performance of the proposed transmission and decoding scheme using Monte Carlo simulations. We focus on comparing it with baseline schemes and theoretical performance upper bounds in scenarios involving massive user access and short-packet coding. Additionally, we verify that as decoding iterations proceed, the simulated decoding error rate converges to the fixed point derived from our analysis. We also confirm the applicability of our coding scheme and fixed-point analysis in random multiple access scenarios. Throughout the evaluations in the first three subsections, we employ rate-$1/3$ regular RA code for channel coding. In the final subsection, we demonstrate the performance of the proposed scheme using different channel codes and conduct comparisons with the performance converse bound.

\subsection{Comparison with Existing Works}
In Figs.~\ref{fig:BER_frozen} and \ref{fig:BER_nonODMA}, we compare BER performance over quasi-static Rayleigh fading channels among the proposed scheme, the scrambling+ML scheme \cite{Yifei}, the case with perfectly known channels, and the non-ODMA (coding+spreading) scheme. We consider $K=300$ communication users with transmission length $n=30000$ and the data length of each user is $B=100$. All four cases adopt the same rate-$R=1/3$ regular RA code as the channel code, so the code length is $L=300$. We considered inserting varying numbers of $P$ frozen bits into the codeword, with their positions randomized. The former three schemes (Fig.~\ref{fig:BER_frozen}) employ the ODMA-based access mechanism, where each user transmits data only in $300+P$ out of the $30000$ time slots, with the remaining slots being idle (empty). In contrast, the fourth scheme (Fig.~\ref{fig:BER_nonODMA}) does not adopt ODMA; instead, it occupies all time slots using spreading. In the simulations, the $E_b/N_0$ and overall transmission rate were fixed, meaning the energy and bandwidth costs of the frozen bits and spreading were accounted for. 

Fig.~\ref{fig:BER_frozen} shows that the scrambling+ML scheme \cite{Yifei} can indeed operate with no frozen bits $(P=0)$, while our proposed scheme, which employs a single-variable LMMSE estimation, requires the assistance of few frozen bits to function. However, inserting a small number of frozen bits improves the performance of both schemes. Moreover, our scheme outperforms the scrambling+ML scheme when the number of frozen bits exceeds 5, and its performance approaches that with perfect channel knowledge. In addition, the scrambling+ML scheme suffers from substantially higher computational complexity.  Notably, when the number of frozen bits exceeds $20$, further increasing their count leads to a gradual performance degradation in decoding, due to the additional energy consumption. In Fig.~\ref{fig:BER_nonODMA}, the performance of non-ODMA scheme suffers severe degradation in massive access scenarios owing to their dense access nature even though both schemes employ the same receiver algorithm.

\subsection{Convergence to Fixed Points}
 In this section, we validate the effectiveness of the proposed fixed-point analysis and characterize the convergence speed of iterative decoding. We adopt the same system parameters as above ($B=100$, $n = 30000$, rate-$1/3$ RA code) and set the number of frozen bits to $P = 20$. To guarantee decoding convergence, we set the maximum number of global iterations for multi-user channel estimation and decoding to 120, and each single-user decoding component operates without inner iterations.
 
\begin{figure}[t]
 	\includegraphics[width=
 	3.3 in]{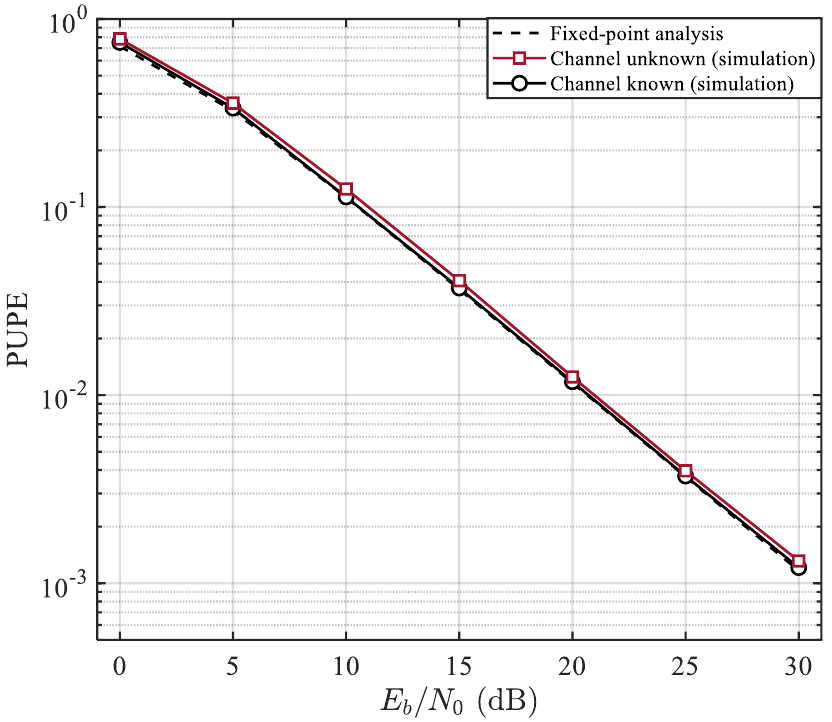}
 	\centering
 	\caption{Comparison between simulated PUPE and fixed-point analysis results with a maximum of 120 global iterations. Each single-user decoding module runs without inner iterations. User number $K = 300$, information data length
 		$B = 100$, transmission length $n = 30000$, number of frozen bits 
 		$P=20$, and rate-$1/3$ regular-RA code for
 		each user. } \label{fig:SimVtheo}
 \end{figure}

\begin{figure}[t]
	\includegraphics[width=
	3.3 in]{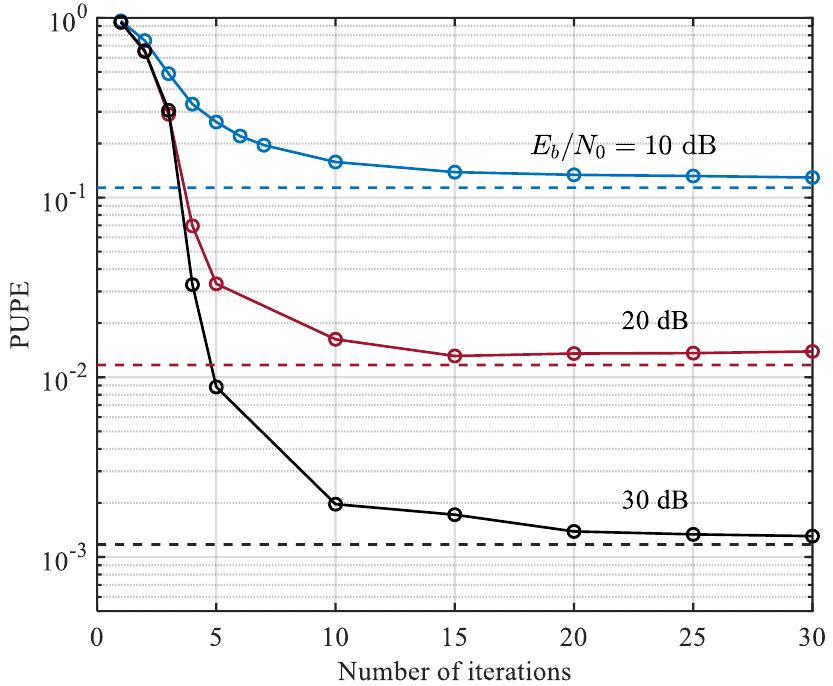}
	\centering
	\caption{Convergence of PUPE with decoding iterations. User number $K = 300$, information data length
		$B = 100$, transmission length $n = 30000$, number of frozen bits 
$P=20$, and rate-$1/3$ regular-RA code for
		each user.} \label{fig:PUPE_iter}
\end{figure}

 As illustrated in Fig.~\ref{fig:SimVtheo}, under the given parameters, our transceiver scheme achieves PUPE close to that with perfect channel knowledge across different $E_\text{b}/N_0$ values, which matches the theoretical performance predicted by the fixed-point analysis. Fig.~\ref{fig:PUPE_iter} reveals that the main performance gain is obtained within the first few iterations. Thus, satisfactory performance can be achieved even with a small number of iterations.
 
\subsection{Performance in Random Multiple Access}
\begin{figure}[t]
	\includegraphics[width=
	3.4 in]{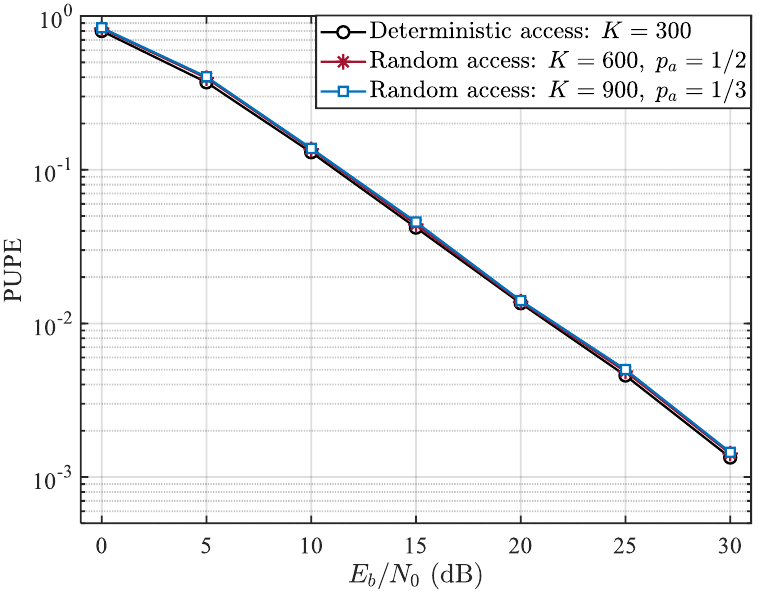}
	\centering
	\caption{Performance of the proposed scheme in random multiple access, where $K$ is the total potential user number and $p_a$ is the user activation probability. Information data length
		$B = 100$, transmission length $n = 30000$, number of frozen bits 
		$P=20$, and rate-$1/3$ regular-RA code for
		each user.} \label{fig:randomaccess}
\end{figure}

Although we have so far only considered the non-random multiple access scenario with a deterministic number of users, our proposed transmission scheme and fixed-point analysis can be directly extended to the random multiple access case. In Fig.~\ref{fig:randomaccess}, we compared the performance of random multiple access system with $600$ and $900$ potential users (with per-user activity probabilities of $p_a=1/2$ and $1/3$, respectively) and the deterministic multiple access system with fixed $300$ active users. All systems adopt identical coding parameter settings ($B=100, n = 30000, P = 20$, rate-$1/3$ RA code) and the same multi-user decoding algorithm. Note that in random multiple access, user activity can be readily determined by checking whether its channel estimate is close to zero. Similar PUPE performance can be observed among the three systems. Therefore, in massive access scenarios, the code design for random multiple access systems can draw upon insights derived from  deterministic multiple access systems.

\subsection{Performance of the Proposed Scheme with Various Channel Codes}
\begin{figure}[t]
	\includegraphics[width=
	3.5 in]{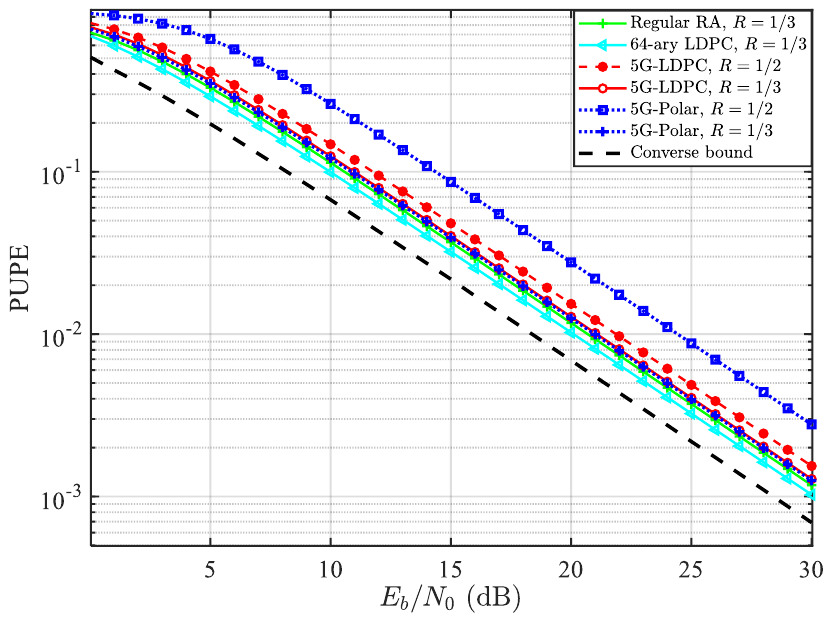}
	\centering
	\caption{Comparison of fixed-point performance for different channel codes with the number of users $K=300$, information data length 
		$B=100$, and transmission length $n=30000$. The single-user code rates are set to $1/2$ and $1/3$.} \label{fig:Various}
\end{figure}
\begin{figure}[t]
	\includegraphics[width=
	3.5 in]{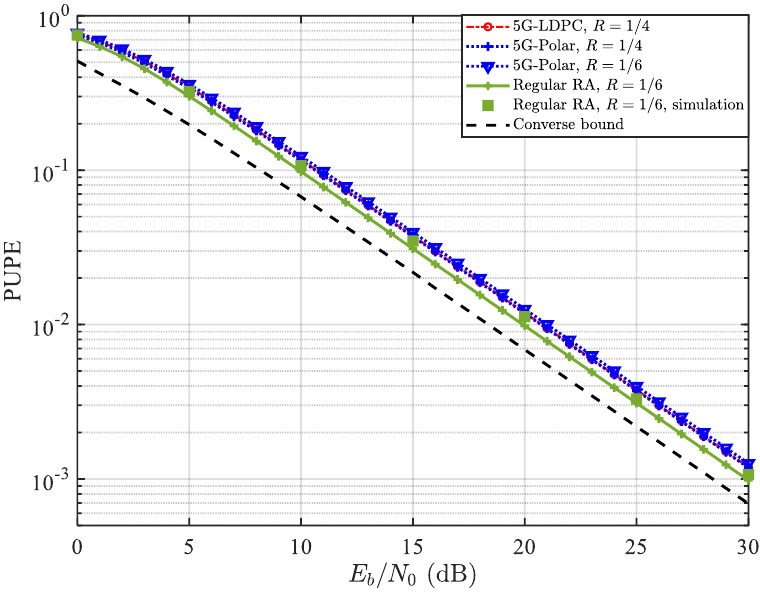}
	\centering
	\caption{Comparison of fixed-point performance for different channel codes with the number of users $K=300$, information data length 
		$B=100$, and transmission length $n=30000$. The single-user code rates are set to $1/4$ and $1/6$.} \label{fig:Various2}
\end{figure}

Figures~\ref{fig:Various} and \ref{fig:Various2} present the fixed-point performance of ODMA employing different channel codes with $K=300$ users, information data length $B=100$, and transmission length $n=30000$. In terms of PUPE performance, low-rate RA code with rate $1/6$ achieves the optimal performance, exhibiting an $E_b/N_0$ gap of approximately 1.5 dB from the theoretical converse bound.
The converse bound is derived as the single-user outage probability for transmitting at an information rate of $B/n$ according to channel capacity, which is formulated as
\begin{equation}
\mathrm{Pr}\left(\log_2\left(1+|h|^2\frac{BE_\text{b}}{nN_0}\right)<\frac{B}{n}\right)=1 - \exp\left( \frac{n N_0}{BE_\text{b}} \left(1-2^{\frac{B}{n}}\right) \right) \label{eq:converse}
\end{equation}
where $h$ denotes the Rayleigh fading coefficient. Since this bound is established based on the single-user outage scenario without considering multi-user interference, it serves as a fundamental lower bound for the PUPE performance of multi-user coding systems. Although the rate-$1/3$ 64-ary LDPC code also delivers excellent PUPE performance, it imposes a significantly higher decoding complexity compared with binary RA codes.

\section{Conclusion}
For massive user access in short-packet transmission scenarios, we proposed a novel ODMA scheme achieving joint multi-user channel estimation and decoding. The core design embedded a small number of frozen bits into codewords and exploited the iterative decoding process to assist channel estimation, which eliminated the requirement of long pilot sequences and spreading operations used in conventional multiple-access systems. Simulation results demonstrated that for a system with 300 users, merely $5\sim 20$ frozen symbols per user were sufficient to approach the performance under perfect channel knowledge. The adopted receiver featured low-complexity modules, including single-user LMMSE channel estimation requiring only correlation operations and iterative interference cancellation.

Furthermore, we developed a fixed-point analysis method for large-scale sparse ODMA systems. The proposed method accurately evaluated the iterative convergence performance of multi-user decoding over block fading channels, solely relying on decoding transfer functions acquired under single-user AWGN channels. It was applicable to performance analysis for channel codes with arbitrary code lengths, code rates and decoding algorithms. Moreover, this analytical framework cut down the enormous time overhead brought by extensive Monte Carlo simulations, and provided an efficient theoretical tool for the optimization and design of multi-user channel codes.

\appendix 
\subsection{Derivation of LMMSE Estimate for \( h_k \)}\label{app:LMMSE}
To simplify the description, we omit the superscripts and subscripts in the notation. Thus, we consider the following model:
\[
\mathbf{y} = h \mathbf{x} + \mathbf{z}
\]
where:
\begin{itemize}
	\item \(\mathbf{x} \) is an \( n \)-dimensional independent Bernoulli random vector, with each element \( x_i \in \{1, -1\} \) and \( E[x_i] = a_i \).
	\item \( \mathbf{z} \) is an \( n \)-dimensional independent complex Gaussian noise vector, with each element \( z_i \) having mean 0 and variance \( V_i \) (i.e., real and imaginary parts each have variance \( V_i/2 \)).
	\item \( h \) is the Rayleigh fading coefficient, modeled as a zero-mean complex Gaussian random variable, \( h \sim \mathcal{CN}(0, 1) \), and independent of \( \mathbf{x} \) and \( \mathbf{z} \).
\end{itemize}

The LMMSE estimate of \( h \) given \( \mathbf{y} \) is given by:
\[
\hat{h} = E[h] + \mathbf{C}_{hy} \mathbf{C}_{yy}^{-1} (\mathbf{y} - E[\mathbf{y}])
\]
Since \( E[h] = 0 \) and \( E[\mathbf{y}] = E[h \mathbf{x} + \mathbf{z}] = 0 \) (due to independence and zero mean), this simplifies to:
\[
\hat{h} = \mathbf{C}_{hy} \mathbf{C}_{yy}^{-1} \mathbf{y}
\]
where \( \mathbf{C}_{hy} \) is the cross-covariance vector between \( h \) and \(\mathbf{y} \), and \( \mathbf{C}_{yy} \) is the auto-covariance matrix of \( \mathbf{y} \).

 \subsubsection{Cross-Covariance \( \mathbf{C}_{hy} \)}
We compute \( \mathbf{C}_{hy} = E[h \mathbf{y}^H] \) (since means are zero). For each element \( y_i = h x_i + z_i \), we have:
\[
E[h y_i^*] = E[h (h^* x_i + z_i^*)] = E[|h|^2 x_i] + E[h z_i^*] = E[x_i] = a_i.
\]
Thus, the cross-covariance vector is:
\[
\mathbf{C}_{hy} = \mathbf{a}^T
\]
where \( \mathbf{a} = [a_1, a_2, \ldots, a_n]^T \).

\subsubsection{Auto-Covariance \( \mathbf{C}_{yy} \)}
The auto-covariance matrix \( \mathbf{C}_{yy} = E[\mathbf{y} \mathbf{y}^H] \) has elements \( E[y_i y_j^*] \). For \( i \) and \( j \):
\[
E[y_i y_j^*] = E[(h x_i + z_i)(h^* x_j + z_j^*)] = E[|h|^2 x_i x_j] + E[z_i z_j^*]
\]
because the cross-terms vanish due to independence. Now:
\begin{align*}
E[|h|^2 x_i x_j] &= E[x_i x_j] \\
E[x_i x_j] &= 
a_i a_j + \delta_{ij} (1 - a_i^2) & \text{since } x_i \in \{1, -1\} \\
E[z_i z_j^*] &= \delta_{ij} V_i
\end{align*}
where \( \delta_{ij} \) is the Kronecker delta. Therefore:
\[
E[y_i y_j^*] =  a_i a_j + \delta_{ij} \left( 1 - a_i^2 + V_i \right)
\]
This can be written in matrix form as:
\[
\mathbf{C}_{yy} = \mathbf{a} \mathbf{a}^T + \mathbf{\Delta}
\]
where \( \mathbf{\Delta} \) is a diagonal matrix with \( i \)-th diagonal element \( d_i = 1 - a_i^2 + V_i \).

\subsubsection{Inversion of \( \mathbf{C}_{yy} \)}
Using the matrix inversion lemma (Woodbury identity):
\[
\mathbf{C}_{yy}^{-1} = \mathbf{\Delta}^{-1} - \frac{ \mathbf{\Delta}^{-1} \mathbf{a} \mathbf{a}^T \mathbf{\Delta}^{-1} }{ 1 +  \mathbf{a}^T \mathbf{\Delta}^{-1} \mathbf{a} }
\]
Let \( c = \mathbf{a}^T \mathbf{\Delta}^{-1} \mathbf{a} = \sum_{i=1}^n \frac{a_i^2}{d_i} \), then:
\[
\mathbf{C}_{yy}^{-1} = \mathbf{\Delta}^{-1} - \frac{  \mathbf{\Delta}^{-1} \mathbf{a} \mathbf{a}^T \mathbf{\Delta}^{-1} }{ 1 + c }
\]

\subsubsection{Final LMMSE Estimate}
Now compute:
\begin{align*}
\mathbf{C}_{hy} \mathbf{C}_{yy}^{-1} &= \mathbf{a}^T \left[ \mathbf{\Delta}^{-1} - \frac{  \mathbf{\Delta}^{-1} \mathbf{a} \mathbf{a}^T \mathbf{\Delta}^{-1} }{ 1 +  c } \right]\nonumber\\
& = \mathbf{a}^T \mathbf{\Delta}^{-1} - \frac{  c \mathbf{a}^T \mathbf{\Delta}^{-1} }{ 1 +  c } \nonumber\\
& = \frac{1}{ 1 +  c } \mathbf{a}^T \mathbf{\Delta}^{-1}.
\end{align*}
Thus, the LMMSE estimate is:
\[
\hat{h} = \frac{1 }{ 1 + \sum_{i=1}^n \frac{a_i^2}{d_i} } \sum_{i=1}^n \frac{a_i y_i}{d_i}
\]
where \( d_i = 1 - a_i^2+ V_i \).

\subsubsection{MSE}
For the LMMSE estimate $\hat{h}$ of $h$ given observations $\mathbf{y}$, the MSE is given by:

\begin{align*}
\text{MSE} &= E[|h - \hat{h}|^2] \\
&= \text{Var}(h) - \mathbf{C}_{hy}\mathbf{C}_{yy}^{-1}\mathbf{C}_{hy}^H\\
&= 1 - \mathbf{a}^T\left(\mathbf{\Delta}^{-1} - \frac{ \mathbf{\Delta}^{-1}\mathbf{a}\mathbf{a}^T\mathbf{\Delta}^{-1}}{1 + c}\right) \mathbf{a} \\
&= 1 - c + \frac{c^2}{1 + c}\\
&=\frac{1}{1 + \sum_{i=1}^n \frac{a_i^2}{d_i}}.
\end{align*}

\subsection{Derivation of Input LLR to DEC}\label{app:LLR}
To simplify the description, we omit the superscripts and subscripts in the notation. Thus, we consider the following model:
\[
y = h x + z,
\]
where:
 \( x \) is a Bernoulli random variable uniformly distributed over \( \{1, -1\} \);
\( h \) is a complex Gaussian random variable with mean \( \hat{h} \) and variance \( e \), i.e., \( h \sim \mathcal{CN}(\hat{h}, e) \); \( z \) is a complex Gaussian random variable with mean 0 and variance \( V \), i.e., \( z \sim \mathcal{CN}(0, V) \).

Given the observation \( y \), the LLR for \( x \) is derived as follows:
\begin{align*}
\text{LLR}(x) &= \log \frac{ \Pr(x=1 \mid y) }{ \Pr(x=-1 \mid y) }\\
 &= \log \frac{ f(y \mid x=1) }{ f(y \mid x=-1) }
\end{align*}
where \( f(y \mid x) \) is the conditional probability density function of \( y \) given \( x \).

Given \( x \), the conditional distribution of \( y \) is complex Gaussian with mean \( x \hat{h} \) and variance \( e + V \) (since \( |x|^2 = 1 \)). Thus,
\[
f(y \mid x) = \frac{1}{\pi (e + V)} \exp\left( -\frac{ |y - x \hat{h}|^2 }{ e + V } \right).
\]
For \( x = 1 \) and \( x = -1 \), the conditional densities are:
\[
f(y \mid x=1) = \frac{1}{\pi (e + V)} \exp\left( -\frac{ |y - \hat{h}|^2 }{ e + V } \right),
\]
\[
f(y \mid x=-1) = \frac{1}{\pi (e + V)} \exp\left( -\frac{ |y + \hat{h}|^2 }{ e + V } \right).
\]
The ratio of these densities is:
\[
\frac{ f(y \mid x=1) }{ f(y \mid x=-1) } = \exp\left( \frac{ -|y - \hat{h}|^2 + |y + \hat{h}|^2 }{ e + V } \right).
\]
Simplifying the exponent:
\[
-|y - \hat{h}|^2 + |y + \hat{h}|^2 = 2(y\hat{h}^* + \hat{h}y^*)=4\Re(y\hat{h}^*).
\]
 Therefore,
\[
\frac{ f(y \mid x=1) }{ f(y \mid x=-1) } = \exp\left( \frac{ 4\Re(y\hat{h}^*) }{ e + V } \right).
\]
Taking the logarithm, the LLR becomes:
\[
\text{LLR}(x) = \frac{ 4\Re(y\hat{h}^*) }{ e + V }.
\]

\end{document}